\documentclass[twocolumn,english,prl,showpacs,floatfix]{revtex4}
\usepackage[T1]{fontenc}
\usepackage[latin1]{inputenc}
\usepackage{graphicx}
\usepackage{amssymb}

\makeatletter

\usepackage{graphicx}
\usepackage{babel}
\makeatother
\begin{document}

\title{On the spin asymmetry of ground states 
in trapped two-component Fermi gases with repulsive interactions}

\author{M. Ögren, K. Kärkkäinen, Y. Yu and S.M. Reimann}

\affiliation{Mathematical Physics, Lund Institute of Technology, P.O. Box 118,
SE-22100 Lund, Sweden}

\date{\today{}}

\begin{abstract}
We examine the spin asymmetry of ground states for 
two-dimensional, harmonically trapped two-component gases of fermionic
atoms at zero temperature with weakly repulsive short range interactions.
Our main result is that, in contrast to the three-dimensional case, 
in two dimensions a non-trivial spin-asymmetric phase can only 
be caused by shell structure.
A simple, qualitative description is given in terms of an approximate
single particle model, comparing well to the standard results 
of Hartree-Fock or direct diagonalization methods.
\end{abstract}

\maketitle

\section{I. Introduction}

Cold atomic gases with multiple components have been realized for example
in Bose-Einstein condensates~\cite{bec,bec2,traprmp1,traprmp2}. 
The nowadays possible experimental studies of atomic Fermi 
gases~\cite{marco,holland,granade,jochim2002,regal2003,greiner2003,zwierlein2003,regal2004,hadzibabic2003}
have raised much recent interest in the different underlying quantum phases, 
with a main attention on pairing.
Fermi gases with spin imbalance have been a hot topic more recently 
in connection with superfluidity 
issues~\cite{zwierlein2006}. While most of the previous studies focus on attractive interactions,  
Duine and MacDonald~\cite{duine2005} pointed out that the repulsive  
regime might offer the possibility of 
ferromagnetism in ultra-cold atom gases with, as they say, 
'unprecedented experimental control'. Describing spin (magnetic) properties of Fermi systems is an important issue 
in many-body physics. The trapped dilute fermionic atom gases, with their
flexibility in tuning important system parameters such as the 
interaction strength, could provide opportunities to look for new 
magnetic phenomena and to test the theoretical models of magnetic 
properties. 

Possible spin
symmetry breaking for a Fermi gas in the unitarity 
limit~\cite{unitarity1,unitarity2,unitarity3,unitarity4,unitarity5} was
discussed by Chevy~\cite{Chevy}.
Maruyama and Bertsch~\cite{marujama2006} have included time-dependent,
external magnetic fields, yielding spin-asymmetric ground 
states.

Sogo and Yabu~\cite{Sogo}  studied the three-dimensional, trapped
Fermi gas with two components. (The more general
multi-component problem was discussed earlier by Salasnich {\it et al.} \cite{salasnich}). These components 
can be characterized by different intrinsic spin states, 
or for example, two hyperfine states, playing the role of
pseudospin in a two-component Fermi gas. 
For weak interactions between the components, the zero-temperature 
ground state is symmetric, with equal up- and down-densities and vanishing 
total spin or pseudospin. Sogo and Yabu showed, 
that in three spatial dimensions, depending on the interaction strength
between the components, a spin-asymmetric, so-called ``collective
ferromagnetic'' state may become energetically favorable.
This spin or pseudospin asymmetry in the ground state 
arises as it becomes energetically favorable 
to increase the Fermi energy of one component 
rather than increasing the interaction energy between the two species. 
Sogo and Yabu~\cite{Sogo} based their investigation on 
the Thomas-Fermi approximation \cite{brack, ring}, an efficient way 
to compute the ground state properties of a large system -- however, 
at the cost of ignoring the shell corrections~\cite{brack}.  

These studies were performed for a three-dimensional harmonic trap. 
However, by appropriate manipulation of the trapping potential,
confined ultra-cold atoms can achieve conditions of reduced
dimensionality. We thus found it useful to further investigate the 
question of spin asymmetry 
for an isotropic harmonic trap in {\it two dimensions}.  
(In one dimension, a uniform fermi system can be solved exactly
with a Bethe ansatz~\cite{Gaudin}. Recently, this was also generalized
to a harmonic trap within the local density approximation~\cite{Hui}).

We found that in two dimensions the spin asymmetry can only arise from
the shell-structure and Hund's rule.
 
\section{II. The Thomas-Fermi approximation for a two-component fermi gas}

Consider fermionic particles of equal mass $m$ at zero temperature, with two 
internal (spin) states labeled by $\sigma \in \{ u,d \}$. 
The fermions shall be subject to repulsive two-body contact 
interactions. 
In atomic units ($\hbar=\omega=m=1$), our model Hamiltonian in two
dimensions is ($r=|\mathbf{{r}}|$)
\begin{equation}
H=\sum_{\sigma} \sum_{i}^{N_{\sigma}}\left[-\frac{1}{2}\Delta_{i}
+\frac{1}{2} r_{i}^{2}\right]+g\sum_{i,u<j,d}\delta\left(\mathbf{{r}}_{i}-\mathbf{{r}}_{j}\right),
\label{hamiltonian}
\end{equation}
where $N_{\sigma}$ denotes the number of particles in each of the two
spin states. Here, only the repulsive, weak-interaction regime of Eq.~({\ref{hamiltonian}) 
is considered. The coupling constant $g\propto a/l_z>0$ is
proportional to the ratio of the three-dimensional scattering length $a$
and the length scale $l_z$ of the ground state in the frozen-out 
$z-$dimension. From the condition 
that the (mean-field) interaction energy should be much smaller than the
Fermi energy $\mu_F$
\begin{equation} 
gn_u^{0}/\mu_{F} \ll 1, 
\end{equation}
with $n_u^{0}$ denoting the particle
density (of one of the species) in the center of the trap, one is led to the diluteness condition 
\begin{equation}
0\leq g\ll2\pi .
\end{equation}
We start by treating the problem within the Thomas Fermi (TF) approximation
which should be valid in the limit of large particle numbers 
where the specific contribution from shell structure plays a minor
role. In the Thomas-Fermi approximation, the total energy is
\begin{equation}
E=\int d\mathbf{{r}}\left[\pi \left( n^2_{u}+  n^2_{d}\right)+
gn_{u}n_{d}+\frac{1}{2}r^{2}
n\right],
\label{Edef}\end{equation}
where
$n_{u }$ and $n_{d }$ and 
$n=n_{u }+n_{d }$ 
are the (pseudo)spin- and total densities, respectively. 
The Euler-Lagrange equations for the densities, while fulfilling the
constraint of constant particle number $N=N_{u}+N_{d}=\int
d\mathbf{{r}}~n ,$ are 
\begin{equation}
\frac{\delta\left(E-\mu_{\sigma} N_{\sigma}\right)}{\delta n_{\sigma}}=0
~{\Rightarrow }~
\left\{ \begin{array}{l}
2\pi n_{u}+gn_{d}+r^{2}/2=\mu_{u}\\
2\pi n_{d}+gn_{u}+r^{2}/2=\mu_{d}\end{array}\right.\label{neqdef}\end{equation}
 In order for the system to be in equilibrium we must have
$\mu_{u}=\mu_{d}\equiv\mu$. Further on for the system
(\ref{neqdef}) to be a stable solution we must have
\begin{equation}
\left|\begin{array}{cc}
\frac{\delta^{2}E}{\delta n_{u}^{2}} & \frac{\delta^{2}E}{\delta n_{u}\delta n_{d}}\\
\frac{\delta^{2}E}{\delta n_{d}\delta n_{u}} & \frac{\delta^{2}E}{\delta n_{d}^{2}}\end{array}\right|\geq0\:\Leftrightarrow\: g\leq 2\pi,\label{detdef}\end{equation}
which is guaranteed by the diluteness condition. By viewing (\ref{neqdef})
as a matrix equation in the variables $n_{u}$ and $n_{d}$
it has the system determinant $4\pi^{2}-g^{2}$. So for $g<2\pi$, which
is again guaranteed by the diluteness condition, the system (\ref{neqdef})
has a unique algebraic solution which corresponds to a stable minimizer
of Eq. (\ref{Edef})
\begin{equation}
\left\{ \begin{array}{l}
n_{u}=\frac{1}{2\pi+g}\left(\mu-\frac{1}{2}r^{2}\right),\: r\leq\sqrt{2\mu}\\
n_{d}=n_{u}\end{array}\right..\label{TFdens}\end{equation}
This TF result suggests that there should be no stable spin-asymmetric
density configurations in the two-dimensional system under study.
This is qualitatively different from the corresponding 
TF investigation for three-dimensional
system where a non-trivial phase separation called ``collective
ferromagnetic state'' has been reported \cite{salasnich, Sogo}. This is due to
the different exponent of the kinetic energy functional in the
three-dimensional case, leading to  non-linear equations for the densities.

Inserting the density (\ref{TFdens}),
$n=n_{u}+n_{d}=2n_{u}$, into the energy functional
(\ref{Edef}), the total energy equals  
\begin{equation}
E =\frac{2}{3}\sqrt{1+\frac{g}{2\pi}}N^{3/2}.
\label{TFtotE}
\end{equation}
We then note that for $g=2\pi$ this energy coincides with the fully
polarized phase, i.e. $1/2$ times Eq. (\ref{TFtotE}) with $g=0$ and $N
\rightarrow 2N$, hence giving a physical motivation for the condition
in Eq. (\ref{detdef}). This describes the trivial phase
separation to a totally polarized system similar to what can happen in three
dimensions \cite{Sogo}.

\section{III. Thomas-Fermi single-particle model}

Let us now incorporate the shell structure. Obviously, the density profile 
given by the above Eq. (\ref{TFdens}) will also be valid for two unequally 
populated non-interacting ($g=0$) spin states, as according to 
Eq.~(\ref{neqdef}), $n_u$ and $n_d$ are then uncoupled. This is the situation 
when particles of one (pseudo)spin state are filling a higher shell than 
the other spin state, such that $|\mu_u-\mu_d|=\omega_{{\it {eff}}}$
  ($\omega_{{\it {eff}}}$ is defined in Eq. (\ref{defomegaeff})). 
The major effect of the 
repulsive interaction is to redistribute atoms from the center of the cloud 
to larger $r$ values. This is qualitatively modeled by the profile of the 
upper Eq. (\ref{TFdens}) also when $|\mu_u-\mu_d|=\omega_{{\it {eff}}}$ (and agrees well with 
a numerical solution of Eq. (\ref{neqdef}) for $g \ll 2\pi$). Moreover, the 
profile of Eq. (\ref{TFdens}) does not break the $SU(2)$ symmetry, as also was shown to be the 
case for the exact density \cite{Zyl}.

We now begin from the single-particle Hartree-Fock equations \cite{brack} for the Hamiltonian
(\ref{hamiltonian}) in the weakly repulsive regime
\begin{equation}
\left[-\frac{1}{2}\Delta+gn_{d}\left(x,y\right)+\frac{1}{2}\left(x^{2}+y^{2}\right)\right]\varphi_{i,
  u}=E_{i, u}\varphi_{i, u}.\label{HFE}\end{equation}
Now, for the purpose of presenting a simple
analytical model, we approximate the particle density with the $g$-dependent TF
density corresponding to Eq. (\ref{TFdens})
\[
n_{d}\left(x,y\right)\approx n^{TF}_{d}\left(|\mathbf{{r}}|\right) \equiv \]
\[
\frac{1}{2\pi+g}\left(\mu_{d}-\frac{1}{2}\left(x^{2}+y^{2}\right)\right)\theta\left(\mu_{d}-\frac{1}{2}\left(x^{2}+y^{2}\right)\right),\]
where $\mu_{d}=\sqrt{\left(2+g/\pi\right)N_{d}}$
and $\theta\left(\cdot\right)$ is Heaviside's step-function. After
separating the wavefunctions $\varphi_{i,u}\left(x,y\right)$ in
Eq. (\ref{HFE}) into the Cartesian coordinates we obtain e.g. for the $x$-dimension

\begin{equation}
\biggl[-\frac{1}{2}\frac{d^{2}}{d x^{2}}+\frac{1}{2}\underbrace{\left(1-\frac{g\theta\left(\mu_{d}-x^{2}/2\right)}{2\pi+g}\right)}_{\omega_{{\it {eff}}}^{2}\left(g,N_{d}\right)}x^{2}\biggr]\psi_{i,u}\left(x\right)=
\end{equation}

\begin{equation}
\biggl[\epsilon_{i,u}-\underbrace{\frac{g\mu_{d}\theta\left(\mu_{d}-x^{2}/2\right)}{2\pi+g}}_{\epsilon_{shift}\left(g,N_{d}\right)}\biggr]\psi_{i,u}\left(x\right).\label{defomegaeff}
\end{equation}

\begin{widetext}

\begin{figure}
\includegraphics[scale=0.75]{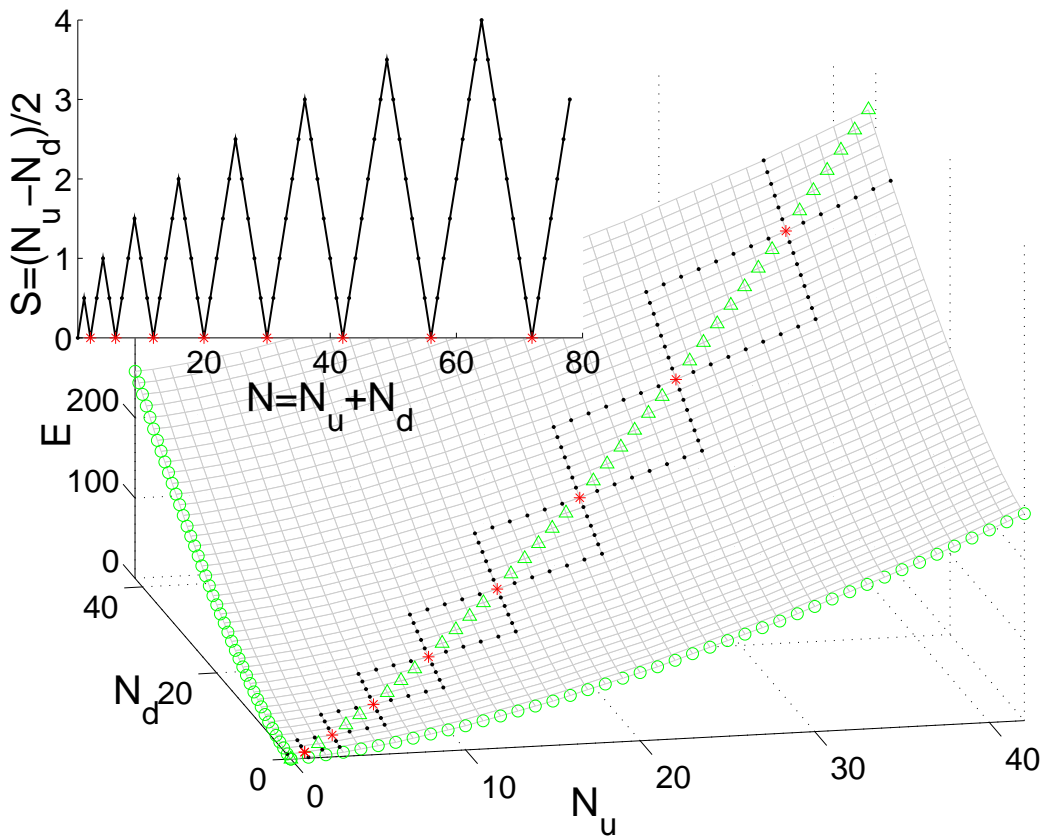}

\caption{(Color online) The 
$N_u$-$N_d$ chart for harmonically trapped 2D fermions
with two spin states according to TFSPM (here with $g=0.1$). The
black dots ($\bullet$) show the states with lowest energy for a given
$N=N_{u}+N_{d}$, the corresponding spin for this
states are seen in the inset figure. The green triangles ($\bigtriangleup$)
show case \emph{ii)} (\emph{spin symmetry}) from equation (\ref{Eexact}),
green circles ($\circ$) show case \emph{iii)} (\emph{polarized}). The
red stars ($\star$) are the magic numbers $M_{j}=j\left(j+1\right)=2,\:6,\:12,\:...$.\label{nuclearchart}}
\end{figure}

\end{widetext}
By re-arranging the terms in Eq. (\ref{defomegaeff}), this can be viewed
as a one-dimensional harmonic oscillator with the perturbation

\begin{equation}
h_p=\left( \frac{g\mu_d}{2\pi+g}-\frac{1}{2}\frac{g}{2\pi+g} x^2\right) \theta\left(\mu_{d}-x^{2}/2 \right).
\end{equation}
We then apply first order non-degenerate perturbation theory in each
Cartesian dimension

\begin{equation}
\epsilon_{i,u}\approx i +\frac{1}{2}+ \langle \phi_i |h_p| \phi_i
\rangle,\label{spenergiesTFSPM}
 \end{equation}
where $\phi_{i}$ is the eigenfunction $i=0,\:1,\:2,\:...$
to the one-dimensional harmonic oscillator with frequency
$\omega=1$. If we now only keep the leading order terms in $g \ll
2\pi$, we can for practical purposes write Eq. (\ref{spenergiesTFSPM}) as
\begin{equation}
\epsilon_{i,u}\approx i +\frac{1}{2}+ \frac{g}{2\pi}
\int_{x^2<\sqrt{8N_d}} \left(\sqrt{2N_d}-\frac{1}{2} x^2  \right)|\phi_i\left( x \right)|^2 dx.\label{integralform}
\end{equation}
Clearly this is the $x$-dependent part of what one obtains if one
treats the mean-field interaction term in Eq. (\ref{HFE}) as $gn_{HO}$ from the beginning,
where $n_{HO}$ refers to the TF-density of the non-interacting system.

Obviously, Eq. (\ref{spenergiesTFSPM}) takes the value for the non-interacting
HO $\epsilon_{i,u}=i+1/2$ in the two limits $g\rightarrow0$
and $N_{d}\rightarrow0$ ($N_{u}\rightarrow N$).
In total,  we have two Cartesian coordinates
and two spin-components such that the total energy is \cite{brack}

\begin{equation}
E_{tot}=E_{u}+E_{d}\approx 2\sum_{occ.}\left(\epsilon_{i,u}+\epsilon_{i,d}\right)-g\int
d \mathbf{{r}}~ n_u n_d.\label{totalenergyTFSPM}
\end{equation}
At zero temperature we have a Fermi-Dirac occupation of each component
such that the HO main quantum number increases by
one unit whenever a shell is filled. There are three special cases
for which the total energy (\ref{totalenergyTFSPM}) can easily be
written down explicitly.\\\emph{i) The non-interacting case:} When $g=0$
we have $\epsilon_{i,\sigma}=i+1/2$.\\\emph{ii) Spin symmetry:}
When $N_{u}=N_{d}$, before  subtracting the last term in Eq. (\ref{totalenergyTFSPM}), the total energy is simply twice
the one for a single component system with the number of particles
being $N/2$ and with a potential $V\left(g,N/2\right)=\omega_{{\it {eff}}}^{2}r^{2}/2+\epsilon_{shift}$,
but in this case the Thomas-Fermi radius $R_{TF}=\sqrt{2\mu_{TF}}$
is equal for both components. Then all the single particle wavefunctions decay
very rapidly for $r>R_{TF}$ and we can formally
drop the factors $\theta\left(\mu_{d}-x^{2}/2\right)$ in
Eq. (\ref{defomegaeff}), then $\epsilon_{i,\sigma}=\omega_{{\it
    eff}}\left( i+1/2\right) +\epsilon_{shift}$.\\\emph{iii)
The totally polarized case:} When $N=N_{u}$ and $N_{d}=0$
(or vice versa) the interaction disappears
which gives $\epsilon_{i,u}=i+1/2$. The cases \emph{i)},
\emph{ii)} and \emph{iii)} are all captured by the formula \cite{MO}

\[
E_{u,d}\left(g,\alpha \right)=\sqrt{\frac{2\pi}{2\pi+g}}\biggl[\frac{1}{6}ceil\left(\frac{\sqrt{1+8\alpha}-1}{2}\right)\times\]

\begin{equation}
\left(6\alpha+1-\left(ceil\left(\frac{\sqrt{1+8\alpha}-1}{2}\right)\right)^{2}\right)+\frac{g}{\sqrt{2}\pi}\alpha^{3/2}\biggr],\label{Eexact}\end{equation}
where $ceil\left(\cdot\right)$ is the closest integer from above.
For case \emph{i)} one has $E_{tot}=E_{u}\left(g=0,\alpha=N_{u}\right)+E_{d}\left(g=0,\alpha=N_{d}\right)$.
In case \emph{ii)} $E_{tot}=2E_{u,d}\left(g,\alpha=N/2\right)-e$
where $e=g\int d \mathbf{{r}} ~n^2/4=gN^{3/2}/\sqrt{36\pi^2+18\pi g}$. In case
\emph{iii)} $E_{tot}=E_{u,d}\left(g=0,\alpha=N\right)$. 

For all other cases, Eq. (\ref{integralform}) can be calculated (numerically) for each single particle level $\epsilon_{i,\sigma}$
which are then summed up according to Eq. (\ref{totalenergyTFSPM}).
If one compares this procedure with a case where Eq. (\ref{Eexact}) is
applicable e.g. case \emph{ii)} where $N_{d}=N_{u}=N/2$,
one finds that the relative difference between those two approaches
is small (e.g. $<2\cdot10^{-3}$ for $g=0.1$ and $N=10$) and decreases with
$N$.

The approximative model we suggest here to illustrate shell-induced
spin asymmetry relies upon the fact that
a spherically symmetric TF particle-density
$n_{TF}\propto\left(\mu-r^{2}/2\right)$ does not break the $SU\left(2\right)$ symmetry, which is present
for the non-interacting problem due to the harmonic trap. It has been
shown in lowest order perturbation theory that the $\ell$-degeneracy
for a given HO shell is not lifted, even when the exact expression
for the particle-density is used \cite{Zyl}. This fact supports the
validity of our approximative ``Thomas-Fermi single particle model'', TFSPM,  
for weak interactions.
However, except for the noninteracting case ($g=0$), the 
TFSPM obviously overestimates the many-body energy 
compared to more sophisticated computational 
methods, where there is more freedom to vary the many-body wavefunction.
TFSPM successfully finds the correct spin-configurations
with lowest energy, see Figure \ref{nuclearchart}, and reproduces
the shell correction with very high accuracy, see Figure \ref{shellcorrections}.
What we here refer to as the correct spin-configuration is what we
obtained from full Hartree-Fock calculations (see below) for all presented values
of $N$. For $N\leq12$ we confirmed the same spin-configuration
by direct numerical diagonalization of the Hamiltonian
(\ref{hamiltonian}).

\section{IV. Comparison to the Hartree-Fock method}

Now we turn our interest to the full Hartree-Fock calculation.
We calculate the particle densities (e.g.) $n_{d}\left(x,y\right)=\sum_{i=1}^{N_{d}}\left|\varphi_{i,d}\right|^{2}$
(without demanding spherical symmetry) for each spin component iteratively
from the mean-field single-particle states $\varphi_{i,d}\left(x,y\right)$
corresponding to the Hartree-Fock Eq. (\ref{HFE}). 

For each total number of particles $N=N_{u}+N_{d}$, we determine 
the lowest-energy state self-consistently. For this state,
the total spin is defined as $S=\left(N_{u}-N_{d}\right)/2$. A
zig-zag pattern with increasing amplitude and frequency occurs, see
inset of Figure \ref{nuclearchart}, which is identical for the Hartree-Fock
calculations and the TFSPM. This pattern (inset of Figure 1) is a
direct consequence of Hund's rule. The trapped atoms at a degenerate
shell can lower their interaction energy by maximizing the number of
atoms of the same species - in other words, by aligning their
spins. This mechanism in contact-interacting fermionic systems leads
to Hund's first rule and what we here call shell-induced spin
asymmetry, in close similarity to long-range interacting
systems. Here, however, Hund's rule has a more dramatic effect since
it removes completely the interaction between atoms of the same spin state.

However, in absolute values, the energy difference associated
with this pattern is a decreasing function of the number of particles.
To be more specific, the lowest energy state carrying the largest
total spin in each shell is $S_{max}\left(j\right)=$$\left(j+1\right)/2$,
where $j=1,\:2,\:3,\:...$ is the index of the shell. Since the magic
numbers of an isotropic two-dimensional harmonic oscillator with two
spin-states are $M_{j}=2\sum_{m=0}^{j-1}\left(m+1\right)=j\left(j+1\right)=2,\:6,\:12,\:...$~,
$S_{max}$ is related to the total number of particles as $S_{max}\simeq\sqrt{N}/2$.
The energy difference of the state with $S\left(N\right)=S_{max}$
compared to the state $S\left(N\right)=0$ is of the order $\Delta E\sim g\sqrt{N}$.
Comparing this to the leading order term in the TF estimate of
the total energy of the state with $S\left(N\right)=0$ from Eq.
(\ref{TFtotE}) $E\simeq2N^{3/2}/3$, the relative energy gain of
the shell-induced spin asymmetry scales as $\Delta E/E\sim g/N$, which becomes
irrelevant for large $N$. This is in agreement with the analysis in
section II, which had shown that
the Thomas-Fermi approximation could not predict any stable spin-asymmetric phase
for the system, but only the trivial, totally polarised system. 

\begin{figure}
\includegraphics[scale=0.47]{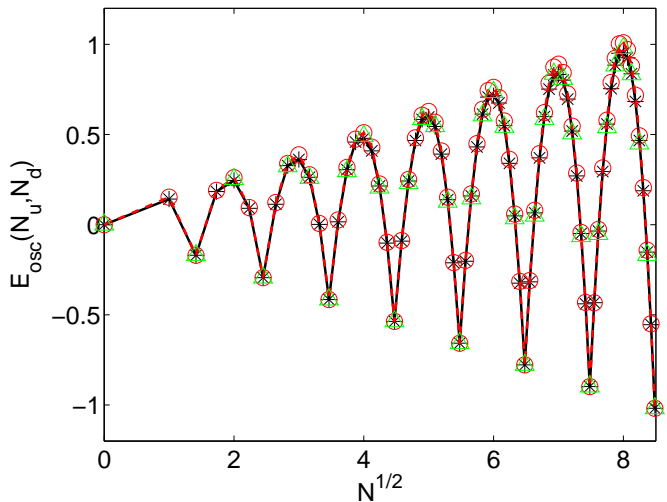}

\caption{(Color online) Oscillating part of the total energy (shell correction)
for the state with lowest energy, at a given $N$ (here with $g=0.1$).
The solid black line with stars ($\star$) shows the TFSPM, 
the red dashed line
with circles ($\circ$) the numerical Hartree-Fock results. 
The green triangles
($\bigtriangleup$) show case \emph{ii)} (\emph{spin symmetry}) from
equation (\ref{Eexact}), which is only defined for even $N$. In this
case the total energy was $\sim 0.3\%$ lower for Hartree-Fock compared
to TFSPM and the
deviations in the shell correction are very small. Investigating
even smaller values of $g$ the curves approach each other further.
The smooth part of the energy is constructed such that the different
curves for the oscillating part should coincide at the minima of the shell
correction (i.e. at the magic numbers) and they follow a line
$y\propto N^{1/2}$.
\label{shellcorrections}}
\end{figure}

\section{V. Conclusions}

In summary, the short-range weakly repulsive interacting fermi gas is a simple
toy model that illustrates very directly Hund's rule: The total spin 
is maximized at mid-shell, in order to minimize the total energy, what
we here have refered to as shell-induced spin asymmetry. 
For the case of an isotropic
two-dimensional harmonic trap we have illustrated a simple approximative
single particle model (TFSPM), that qualitatively reproduces the energy
landscape of the system, when compared to Hartree-Fock calculations.
This model is restricted to two dimensions and weak interactions since
it relies upon the fact that the TF density profile does not break the
$SU\left(2\right)$ symmetry of the non-interacting problem and the
fact that perturbation theory is used. This
also outrules the possibility to see super-shell structure, reported in the
three-dimensional case \cite{Yu}, within this approach.

\section{Acknowledgements}

This work was financially supported by the Swedish Research Council and
the Swedish Foundation for Strategic Research, as well as
the European Community project ULTRA-1D 
(NMP4-CT-2003-505457).


\begin{thebibliography}{1}

\bibitem{bec} M.H. Anderson, J.R. Ensher, M. R. Matthews, C.E. Wieman and
  E.A. Cornell, Science {\bf 269}, 189 (1995).


\bibitem{bec2} K.B. Davis, M.O. Mewes, M.R. Andrews, N.J. van Druten, D.S. Drufee, D.M. Kurn
  and W. Ketterle, Phys. Rev. Lett. {\bf 75}, 3969 (1995).

\bibitem{traprmp1} W. Ketterle, Rev. Mod. Phys. {\bf 74}, 1131 (2002).

\bibitem{traprmp2} E. A. Cornell and C. E. Wieman, Rev. Mod. Phys. {\bf 74}, 875 (2002).

\bibitem{marco} B. De Marco and D.S. Jin, Science {\bf 285}, 1703 (1999).

\bibitem{holland} M.J. Holland, B. DeMarco, and D.S. Jin, Phys. Rev. A {\bf
  61}, 053610 (2000).

\bibitem{granade} S.R. Granade, N.E. Gehm, K.M. O'Hara, and J.E. Thomas, 
Phys. Rev. Lett. {\bf 88}, 120405 (2002).

\bibitem{jochim2002} S. Jochim, M. Bartenstein, G. Hendl, J. Hecker Denschlag, 
and R. Grimm, Phys. Rev. Lett. {\bf 89}, 273202 (2002).

\bibitem{regal2003} C.A. Regal and D.S. Jin, Phys. Rev. Lett. {\bf 90},
230404 (2003).

\bibitem{hadzibabic2003} Z. Hadzibabic, S. Gupta, C. A. Stan, C. H. Schunck,
M. W. Zwierlein, K. Dieckmann, and W. Ketterle, Phys. Rev. Lett. {\bf 91},
160401 (2003).

\bibitem{greiner2003} M. Greiner, C.A. Regal and D.S. Jin, Nature (London)
{\bf 426}, 537 (2003).

\bibitem{zwierlein2003} M. W. Zwierlein, C. A. Stan, C. H. Schunck, 
S. M. F. Raupach, S. Gupta,
Z. Hadzibabic, and W. Ketterle, Phys. Rev. Lett. {\bf 91}, 250401 (2003).

\bibitem{regal2004} C.A. Regal, M. Greiner and D.S. Jin, 
Phys. Rev. Lett. {\bf 92}, 040403 (2004).

\bibitem{zwierlein2006} M. W. Zwierlein, A. Schirotzek, C.H. Schunck, and
W. Ketterle, Science {\bf 311}, 492 (2006).

\bibitem{duine2005} R.A. Duine and A.H. MacDonald, Phys. Rev. Lett. {\bf 95},
  230403 (2005).

\bibitem{unitarity1} ``{\it The Many-Body Challenge Problem}'' formulated by
G. F. Bertsch, 1999, see R. F. Bishop, Int. J. Mod. Phys. B {\bf 15},
 iii (2001).

\bibitem{unitarity2} K. M. O'Hara et al., Science 298, 2179 (2002); 

\bibitem{unitarity3} J. Kinast, A. Turlapov, J. E. Thomas, Q. Chen, J. Stajic,
  and K. Levin, Science {\bf 307}, 1296 (2005); 

\bibitem{unitarity4} T.-L. Ho, Phys. Rev. Lett. {\bf 92}, 090402 (2004); 

\bibitem{unitarity5} A. Bulgac, J.E. Drut, and P. Magierski, 
Phys. Rev. Lett. {\bf 96}, 090404 (2006). 

\bibitem{Chevy}F. Chevy, ``{\it Universal phase diagram of a strongly
  interacting Fermi gas with unbalanced spin populations''}, cond-mat/0605751 (2006).

\bibitem{marujama2006} T. Marujama and G.F. Bertsch, Phys. Rev. A {\bf 73},
013610 (2006).

\bibitem{Sogo}T. Sogo and H. Yabu, Phys. Rev. A \textbf{66}, 043611
(2002).

\bibitem{salasnich} L. Salasnich, B. Pozzi, A. Parola and L. Reatto,
  J. Phys. B: At. Mol. Opt. Phys. {\bf 33}, 3943 (2000).

\bibitem{brack} M. Brack and R. K. Bhaduri: Semiclassical Physics (Westview
  Press, Boulder, USA, 2003).

\bibitem{ring} P. Ring and P. Schuck, The Nuclear Many-Body Problem, 
(Springer-Verlag, New York, 1980).  

\bibitem{Gaudin} M. Gaudin, Phys. Lett. \textbf{24A}, 55 (1967).

\bibitem{Hui} H. Hui, L. Xia-Ji and P.D. Drummond, ``{\it FFLO phase in one-dimensional polarized Fermi gases''}, cond-mat/0610448.

\bibitem{Zyl}B. P. van Zyl, R.K. Bhaduri, A. Suzuki, and M. Brack, Phys. Rev. A \textbf{67}, 023609 (2003).

\bibitem{MO}M. Ögren, unpublished (2004).

\bibitem{Yu}Y. Yu, M. \"Ogren, S. {\AA}berg, S.M. Reimann and M. Brack, Phys. Rev. A \textbf{72}, 051602(R) (2005).

\end{thebibliography}
\end{document}